\documentclass[11pt,twoside]{article}
\usepackage{asp2010}

\resetcounters

\begin{document}

\title{Synthetic Light Curves for Born Again Events: Preliminary Results}
\author{Marcelo M. Miller Bertolami$^{1,2,3}$\& Ren\'e D. Rohrmann,$^{2,4}$
\affil{$^1$Facultad de Ciencias Astron\'omicas y Geof\'isicas, Universidad Nacional de La Plata, Paseo del Bosque s/n, 1900 La Plata, Argentina}
\affil{$^2$Member of CONICET, Argentina}
\affil{$^3$Max-Planck-Institut f\"ur
  Astrophysik, Karl-Schwarzschild-Str. 1, 85748, Garching,
  Germany.}
\affil{$^4$Instituto de Ciencias Astron\'omicas, de la Tierra y del Espacio (CONICET-UNSJ), Av. Espa\~na 1512 (sur), 5400 San Juan, Argentina}}

\begin{abstract}
The development of surveys which will be able to cover a large region
of the sky several times per year will allow the massive detection of
transient events taking place in timescales of years.  In addition,
the projected full digitalization of the Harvard plate collection will
open a new window to identify slow transients taking place in
timescales of centuries. In particular, these projects will allow the
detection of stars undergoing slow eruptions as those expected during
late helium flashes in the post-AGB evolution. In order to identify
those transients which correspond with late helium flashes the
development of synthetic light curves of those events is mandatory. In
this connection we present preliminary results of a project aimed at
computing grids of theoretical light curves of born again stars.
\end{abstract}

\section{Introduction}
About 20\% of the stars leaving the Asymptotic Giant Branch (AGB) are
supposed to undergo a final thermal pulse during their post-AGB
evolution \citep{1984ApJ...277..333I}. In those cases, the star is
predicted to be temporarily reborn as a yellow giant
\citep{1979A&A....79..108S} in the so-called “Born Again AGB” scenario
\citep{1984ApJ...277..333I}. The transition from the prewhite dwarf to
the giant configuration is expected to be very rapid; being of a few
years in the very late thermal pulse (VLTP) case
\citep{1995LNP...443...48I} and of the order of a century in the case
of a late thermal pulse (LTP) \citep{1979A&A....79..108S}. Due to the
short duration of these events, although $\sim 10$\% of all post-AGB
stars are expected to undergo a VLTP, the observation of those events
is very rare. Indeed, only two objects (V605 Aql and V4334
Sgr)\footnote{In addition two other eruptions have been suggested to
  be VLTP events; CK Vul \citep{2002MNRAS.332L..35E} and more recently NSV 11749
  \citep{2011ApJ...743L..33M}, but neither of these have been
  confirmed.}
have been identified as stars undergoing a VLTP
\citep{2000AJ....119.2360D, 2002Ap&SS.279..183D}, while a third has
been identified with the LTP flavor of the scenario (FG Sge, see
\citealt{2006A&A...459..885J} and references therein). Individual born
again stars are very valuable to test our understanding of the
s-processes during the thermal pulses on the AGB
\citep{1999A&A...343..507A, 2006A&A...459..885J} and also to test and
validate the simulations of convective reactive burning in the
interior of the stars \citep{2011ApJ...727...89H}. In addition, these
events are key to understand the formation of carbon-rich H-deficient
stars and H-deficient white dwarfs \citep{2006PASP..118..183W}.
Consequently, the identification of more stars undergoing late helium
shell flashes will boost our understanding of the late phases of low
and intermediate mass stars and their nucleosynthesis. In this
connection, a rough estimate suggests that the birth rate of planetary
nebulae in our galaxy is of about one every year
\citep{2002Ap&SS.279..171Z}. If 10\% of their central stars become
VLTP giants, then we should expect such events in our galaxy to take
place at a rate of about one per decade. Due to their high intrinsic
brightnesses ($M_{\rm V}$=-2...-4) these objects can be easily
detected at large distances within our galaxy. In this context, the
development of surveys which will be able to cover a large region of
the sky several times per year, like the Panoramic Survey Telescope \&
Rapid Response System (PanSTARRS, \citealt{2012ApJ...750...99T}) and
the Large Synoptic Survey Telescope (LSST,
\citealt{2012IAUS..285..158W}), will allow the systematic detection of
born again events.  In addition, the projected full digitalization of
the Harvard plate collection (Digital Access to a Sky Century at
Harvard, DASCH, \citealt{2009ASPC..410..101G}) will open a new window
to identify slower born again events taking place in timescales of
centuries like FG Sge. Due to the lack of enough observational
counterparts of bonafide born again stars the identifications of such
events has to rely partially in our present theoretical knowledge of
their evolution. We have recently started a project to compute, and
provide, theoretical light curves for born again events that will help
in the identification of those transients. In what follows we present
some preliminary results and discuss future developments.

\section{Numerical modeling}
In order to compute synthetic light curves appropriate for Born Again
AGB stars proper stellar evolution models and atmospheres are
needed. As a first step we computed the evolution of stellar models
during fast born again events (low mass VLTP events). During VLTP
events the hydrogen envelope remaining on the hot white dwarf is
convectively mixed into the interior of the star and violently burned
---H-flash, see \cite{2006A&A...449..313M}. Stellar evolution
sequences were computed with {\tt LPCODE} stellar evolution code which
is perfectly suited to compute the mixing \& burning during the
violent H-flash \citep{2005A&A...435..631A}.  Initial models for a
wide range of hot-white dwarf masses were taken from
\cite{2007MNRAS.380..763M}, which is the only available grid of
detailed pre-born again models to date. The masses of the models
studied in the present work are 0.530M$_\odot$, 0.542M$_\odot$,
0.561M$_\odot$, 0.564M$_\odot$, 0.584M$_\odot$ and 0.609M$_\odot$. As
mentioned above, in order to compute appropriate light curves for
different filters, a proper assesment of the emergent spectral energy
distribution of the models is necessary. Synthetic magnitudes for the
stellar evolution models are computed with a modified version of the
stellar atmosphere code described in Rohrmann et al. (2011) which
includes a simplified treatment of the effects of C, N and O in the
structure of the atmosphere and radiative transfer equations. In
brief, the modelling of the stellar atmosphere includes several ions
and molecules of H and He (see \citealt{2011MNRAS.411..781R} for a
detailed description) and the neutral and first three ionization
stages of C, N and O. The chemical balance of
these species is determined by the occupation probability formalism
for H and He \citep{1988ApJ...331..794H} and by the ideal partition
functions for the C, N and O ions. Opacity sources for H and He ions
and molecules are included in a very detailed and exhaustive way
\citep{2011MNRAS.411..781R} while C, N and O ion opacity sources are
included within the hydrogenic approximation and no molecules have
been included so far.

\section{Preliminary results and future work}

\begin{figure}[]
\includegraphics[clip, angle=0, width=10cm]{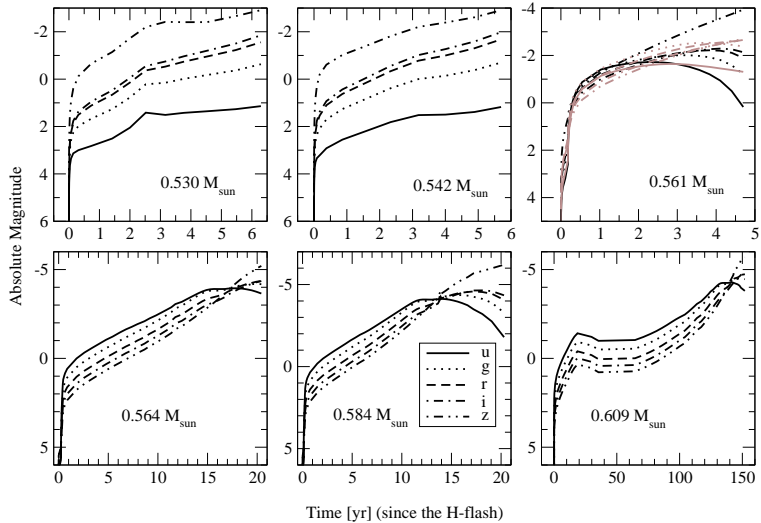} 
\caption{Synthetic light curves computed for the SDSS photometric
  system under the assumption of V4334 Sgr-like compositions for all
  our computed stellar evolution sequences. The effect of different
  compositions is shown in the top right panel for the 0.561M$_\odot$
  where the light curves for a solar-like atmosphere are also shown
  (grey curves). Note that the much slower eruption light curve of the
  0.609M$_\odot$ sequence is due to the fact that massive remnants do
  not harbour enough hydrogen for the H-flash to drive a fast
  expansion as already explained by \cite{2007MNRAS.380..763M}.}
\label{Figure}
\end{figure}

As expected, eruption light curves are dependent not only on the
timescales set by the evolution of the stellar interior, but also on
the specific element abundances of the photosphere of the models.  In
Fig. \ref{Figure} (top right panel) we show the predicted light curves
obtained from the same stellar evolution sequence when different
surface abundances are adopted in the computation of the synthetic
magnitudes. Preliminary synthetic light curves were computed for the
UBVRJ \citep{1990PASP..102.1181B} and the Sloan Digital Sky Survey
(ugriz; \citealt{1996AJ....111.1748F}) photometric systems. In
Fig. \ref{Figure} we show the computed light curves for all our
stellar evolution sequences in the SDSS photometric system, when
abundances similar to those observed in V4334 Sgr
\citep{1999A&A...343..507A} are assumed. Note that the SDSS
photometric system is similar to the one adopted by the PanSTARRS1
survey \citep{2012ApJ...750...99T}.  In order to assess the importance
of the detailed treatment of the atmosphere and also the importance of
the different atmospheric abundances, two test cases have been
computed by assuming different atmospheric compositions for the
0.561M$_\odot$ sequence: a solar-like atmosphere ($X_{\rm H}/X_{\rm
  He}/X_{\rm C}/X_{\rm N}/X_{\rm O}/= 0.74/0.25/2.4\times 10^{-3}/
7\times 10^{-4}/ 5.8\times 10^{-3}$, brown lines) and a V4334 Sgr-like
atmosphere ($X_{\rm H}/X_{\rm He}/X_{\rm C}/X_{\rm N}/X_{\rm O}/=
7.4\times 10^{-4}/ 0.74/0.22/8.2\times 10^{-3}/ 2.9\times 10^{-2}$,
black lines, \citealt{1999A&A...343..507A}). Clearly, a correct
assessment of the radiative transfer in the atmosphere is necessary to
compute born again light curves.

Several improvements in theoretical lightcurves, as well as a proper
assessment of the involved uncertainties, are still necessary. In
particular we plan to improve the stellar atmospheric modelling by
including a better treatment of the C, N and O ions and the inclusion
of related molecules. Also a calibration of the mixing efficiency in
stellar evolution sequences during the violent H-flash is necessary
and will be performed. In addition we plan to expand the grid of
masses to include higher masses, with slower VLTP events, and to compute
lightcurves for LTP episodes.

\acknowledgements M3B thanks the organizers of the EUROWD12 for the
finantial assistance that helped him to attend the conference. This
research was supported by PIP 112-200801-00940 from CONICET and
PICT-2010-0861 from ANCyT.

\bibliographystyle{asp2010}
\bibliography{M3B_RR}

\end{document}